\documentclass{article}
\usepackage{ijcai19}

\usepackage{xcolor}

\usepackage{times}
\usepackage{soul}
\usepackage{url}
\usepackage[hidelinks]{hyperref}
\usepackage[utf8]{inputenc}
\usepackage[small]{caption}
\usepackage{graphicx}
\usepackage{subfigure}
\usepackage{amsmath}
\usepackage{amsfonts}
\usepackage{booktabs}
\usepackage{algorithm}
\usepackage{algorithmic}
\usepackage{epstopdf}

\newcommand{\eat}[1]{}

\newtheorem{definition}{Definition}

\title{Revisiting Flow Information for Traffic Prediction}

\eat{
}

\author{
	Xian Zhou
	\and
	Yanyan Shen\And
	Linpeng Huang
	\affiliations
	Department of Computer Science and Engineering\\
	Shanghai Jiao Tong University\\
	\emails
	\{zhouxian, shenyy, lphuang\}@sjtu.edu.cn
}

\begin{document}

\maketitle

\begin{abstract}

Traffic prediction is a fundamental task in many real applications, which aims to predict the future traffic volume in any region of a city. In essence, traffic volume in a region is the aggregation of traffic flows from/to the region. However, existing traffic prediction methods focus on modeling complex spatiotemporal traffic correlations and seldomly study the influence of the original traffic flows among regions. In this paper, we revisit the traffic flow information and exploit the direct flow correlations among regions towards more accurate traffic prediction. We introduce a novel flow-aware graph convolution to model dynamic flow correlations among regions. We further introduce an integrated Gated Recurrent Unit network to incorporate flow correlations with spatiotemporal modeling. The experimental results on real-world traffic datasets validate the effectiveness of the proposed method, especially on the traffic conditions with a great change on flows.

\end{abstract}

\section{Introduction}

\begin{figure}[t]
	\vspace{.1in}
	\centering
	\includegraphics[width=0.9\linewidth]{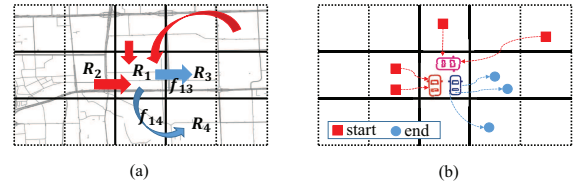}
	\vspace{-.05in}
	\caption{
		Illustration of traffic volume and flow. 
		For example, the out-flows of $R_1$ is summation of traffic flow $f_{13} = 2$ to $R_3$ and $f_{14} = 1$ to $R_4$, i.e., 3. 
	}
	\label{fig:intro-flow}
	\vspace{-.1in}
\end{figure}

Predicting traffic volume in any region of a city has become one of most fundamental problems in nowadays intelligent transportation systems~\cite{ijcai/YuYZ18}. Typically, \emph{traffic volume} of a region is defined as a two-dimensional vector, measuring the total numbers of traffic in-flows and out-flows caused by mobility trips (e.g., taxi trajectories), respectively. For example, as shown in Figure~\ref{fig:intro-flow}, there are $3\times5=15$ regions. During time $t$, $4$ out of $7$ mobility trips (in red) ended in region $R_1$ and the remaining $3$ trips (in blue) started from $R_1$. Hence, the traffic volume of $R_1$ at time $t$ is ($4,3$), representing the in-flows and the out-flows of $R_1$ at that time. 

Various approaches have been proposed to predict citywide traffic volume based on historical traffic data. Conventional methods such as ARIMA and its variants~\cite{Moreira2013Predicting,Lippi2013Short} considered the traffic volume in a region over time as a univariate time series and learned a regression function for traffic prediction under the simple stationary stochastic process assumption. Some works~\cite{gis/LiZZC15,kdd/TongCZCWYYL17} incorporated spatial features or external factors to enhance the information of each region and leveraged machine learning models for traffic prediction. 
To improve the prediction accuracy, many recent studies~\cite{aaai/ZhangZQ17,yao2018deep,yao2019revisiting} have focused on developing deep neural networks for modeling complex and dynamic \emph{spatio-temporal traffic correlations} among regions.
The rationale behind is that traffic volume of a region is dependent of that of its surrounding regions (i.e., spatial correlation) as well as its volume in previous time periods (i.e., temporal correlation).
They thus combined convolutional neural networks (CNN) and recurrent neural networks (RNN) to jointly learn deep spatio-temporal features from previous traffic volume of regions, and achieved the state-of-the-art performance. 

All the existing works have devoted great effort to exploiting informative features and their deep interactions from historical traffic volume for future traffic prediction. We notice that traffic volume of a region is the aggregation result of traffic flows from/to other regions. In addition to leveraging past traffic volume for prediction, can we move one step back and realize the influence of original traffic flows among regions on future traffic volume? 

Intuitively, modeling traffic flow correlations among regions is a more direct and precise way to capture the actual traffic correlations. For example, a large quantity of traffic flows from region A to region B at some time period may cause 
(i) both increases in the out-flows of A and the in-flows of B,
or (ii) a reverse flow direction from B to A in the future due to the return trips. 
Being aware of such \emph{flow correlations} is undoubtedly beneficial to enhance traffic prediction performance. 
Nevertheless, existing CNN based models are incapable of capturing direct flow correlations among regions. 
The reasons are two-fold. 
First, the two regions A and B with traffic flow may not be close to each other geographically, and the convolutional filters with limited sizes can only focus on a small number of nearby regions.
Second, the direct flow correlations among regions are dynamically changing, but CNNs are known to have geographically fixed filters~\cite{iccv/DaiQXLZHW17,nips/JiaBTG16}. 
That is, the learned convolutional filters will assign the same weight to a region over time due to its fixed filtering mechanism, which fails to model the flow dynamics even for nearby regions.
%


To this end, this paper introduces a novel flow-aware traffic prediction model which aims to exploit the direct flow correlations among regions for better prediction performance.
We organize the original traffic flows into a graph structure where each vertex in the graph corresponds to a region and the directed edge between two vertices are weighted by the dynamic flows. A novel attempt of our approach is a \emph{flow-aware graph convolution} module to reveal the traffic correlations among regions based on the flow graphs. Specifically, for each region $r$, the graph convolution defines a \emph{dynamic receptive field} that automatically recognizes a set of regions that are highly correlated with $r$.  It is worth mentioning that (i) the identified correlated regions may not be geographically close to $r$, and (ii) the set of regions will dynamically change according to the actual traffic flows. 
%
We also apply the typical convolution to capture the spatial correlations for nearby regions. An integrated Gated Recurrent Unit (GRU) network is further applied to absorb the outputs from two kinds of convolutions and learn the traffic tendency effectively. 
We enforce an early combination of the features extracted from two convolutions at each time step to model deep feature interactions. We dub our proposed integrated block as \textbf{FlowConvGRU}. Finally, we stack multiple FlowConvGRUs into a deep structure, followed by a fully connected layer to predict the future traffic. 
%
%
%
We evaluate the performance of our proposed approach using two real-world traffic datasets.
Extensive experimental results demonstrate that our proposed method achieves higher prediction accuracy than various state-of-the-art methods and the flow-aware graph convolution can obtain deeper insights over flow information towards better prediction performance.

\section{Preliminaries}\label{sec:prelim}

\subsection{Definitions}\label{sec:def}

Suppose a city area is divided into $N=m\times k$ disjoint regions with equal sizes. Let $\cal{T}$ denote the set of mobility trips, where $T=(i,j,t_s, t_e)\in \cal{T}$ represents a trip starts from region $i$ at time $t_s$ and ends at region $j$ during time $t_e$. Note that $t_s \leq t_e$.

\begin{definition}[Traffic flow $f_{ij}^t$]\label{def:flow}
	Given $\cal T$, the traffic flow $f_{ij}^t$ from region $i$ to region $j$ at time $t$ can be computed by:
	$$f_{ij}^t = | \{ T\in{\cal T} \mid T.i = i \wedge T.j = j \wedge T.t_s \leq t \wedge T.t_e = t  \}|$$
	Let $f^t = \{f^t_{ij}\}$ be the traffic flow matrix at time $t$.
	Specifically, $f^t_{:i}=\{f^t_{ji}, j=1,\cdots, N\}$ and $f^t_{i:}=\{f^t_{ij}, j=1,\cdots,N\}$ indicate in-flow and out-flow information for region $i$ at time $t$, respectively. 
\end{definition}

\begin{definition}[Traffic flow graph $G^t$]\label{def:flowgraph}
	%
	Given $\cal T$ and $N$ regions, we construct a traffic flow graph at time $t$, denoted by $G^t = (V^t, E^t)$, where $V^t$ is the set of $N$ vertices representing all the regions and $E^t$ is the set of weighted directed edges representing the non-zero traffic flow between two regions at time $t$, i.e., $(i,j)\in E^t$ iff $f_{ij}^t>0$.  
	The weight of each edge $(i,j)$ in $E^t$ is traffic flow $f_{ij}^t$ from region $i$ to region $j$. 
	Note that $f^t$ is actually the weighted adjacency matrix for $G^t$.
\end{definition}

\begin{definition}[Traffic volume of a region $x_i^t$]\label{def:volume}
	Given $\cal T$, the traffic volume $x_i^t$ of a region $i$ at time $t$ includes total numbers of in-flows $\overrightarrow{x_i^t}$ and out-flows $\overleftarrow{x_i^t}$ during time $t$, which can be computed by:
	$$x_i^t=(\overrightarrow{x_i^t}, \overleftarrow{x_i^t}) = (\sum_{k=1}^{N}{f_{ki}^t}, \sum_{k=1}^{N}{|\{T\in{\cal T} \mid T.i=i \wedge T.t_s=t\}|})$$
\end{definition}

We organize citywide traffic volume into a 3D tensor $X^t \in \mathbb{R}^{m\times k \times 2}$, where the last dimension corresponds to traffic volume in each region (i.e., in-flows and out-flows).


\begin{definition}[Traffic prediction]\label{def:problem}
	Given the traffic flow graphs $\{G^1, \cdots, G^T\}$ and traffic volume tensors $\{X^1, \cdots, X^T\}$ in previous $T$ time steps, we focus on predicting the traffic volume tensor in time $T+1$, i.e., $X^{T+1}$.
\end{definition}

\subsection{Graph Convolution Networks}
Graph convolutional networks are proposed to apply convolution operations on graph data, which cannot be handled using typical convolutional networks.
Generally there are two kinds of graph convolution networks.
The \emph{spatial graph convolution networks}~\cite{icml/NiepertAK16} try to rearrange neighboring vertices to match a grid form such that the typical convolution can be applied.
The \emph{spectral graph convolution networks}~\cite{Bruna2014Spectral} implement convolution in the spectrum domain based on graph Fourier transformation.
Some works~\cite{nips/DefferrardBV16,kipf2017semi} developed fast localized convolutional filters on graphs using Chebyshev polynomial parametrization, which consider spatial localization on graph vertices and substantially reduce the computational complexity.

%
Consider an undirected graph $G = (V, E)$ with $n$ vertices.
Suppose the adjacency matrix is $A \in \mathbb{R}^{n\times n}$ and the diagonal degree matrix is $D$, i.e., $D_{ii} = \sum_j A_{ij}$. 
We can calculate the corresponding normalized graph Laplacian as: $L=I_n - D^{-\frac{1}{2}} A D^{-\frac{1}{2}} = U \Lambda U^T$, where $U$ and $\Lambda$ are matrix of eigenvectors and diagonal matrix of eigenvalues, respectively.
The spectral convolution is defined as the multiplication of a graph signal $s \in \mathbb{R}^n$ with a filter $g_\theta (\Lambda)$ which is a diagonal matrix parametrized by $\theta \in \mathbb{R}^n$. Formally, we have:
\begin{equation}
	g_\theta \star s = U g_\theta (\Lambda) U^T s
\end{equation}
where $U^T s$ and $U s$ are the graph Fourier transform and inverse transform of graph signal $s$, respectively.
In~\cite{kipf2017semi} and~\cite{nips/DefferrardBV16}, the authors used a truncated expansion of Chebyshev polynomials $T_k(\cdot)$ to approximate $g_\theta (\Lambda)$ up to $K$-order: $g_{\theta '} (\Lambda) \approx \sum_{k=0}^K \theta_k' T_k (\tilde{\Lambda})$, where $\tilde{\Lambda} = \frac{2}{\lambda_{max}} \Lambda - I_n$ and $\lambda_{max}$ is the largest eigenvalue of $L$.
In this way, the convolution can be reformulated as: 
\begin{equation}
	g_{\theta'} \star s \approx \sum_{k=0}^{K-1} \theta_k' T_k (\tilde{L}) s,
\end{equation}
where $\tilde{L} = \frac{2}{\lambda_{max}} L - I_n$. The approximation achieves $K$-localized convolution and reduces the complexity to $\mathcal{O}(|E|)$.

However, the above spectral graph convolutions require the graph to be undirected to perform matrix decomposition.
Recently, the authors in~\cite{li2018dcrnn_traffic} proposed a diffusion convolution operation for directed graphs.
They represent directed edges as bidirectional diffusion process, 
which is characterized by random walks with
two state transition matrices $D_I^{-1} A^T$ and $D_O^{-1} A$, where $A$ denotes the weighted adjacency matrix of the graph, and
$D_I$ and $D_O$ are the in-degree and out-degree diagonal matrices, respectively.
Similarly, a finite $K$-step truncation of diffusion process is utilized for fast localization.
Formally, the diffusion convolution over the graph signal $s \in \mathbb{R}^n$ with a filter $\theta$ is formulated as:
\begin{equation}
	\theta \star_\mathcal{D} s = \sum_{k=0}^{K-1} (\theta_{k,1} (D_O^{-1}A)^k + \theta_{k,2} (D_I^{-1} A^T)^k) s
\end{equation}
where $K$ is the diffusion step and $\star_\mathcal{D}$ represents the diffusion convolution. $\theta \in \mathbb{R}^{K\times 2}$ is the learned parameters for the filter. 
Since traffic flow is expected to be directed among regions, we leverage diffusion convolution to model flow correlations for traffic prediction.
\eat{
\begin{figure*}[t]
	\centering
	\includegraphics[width=0.9\linewidth]{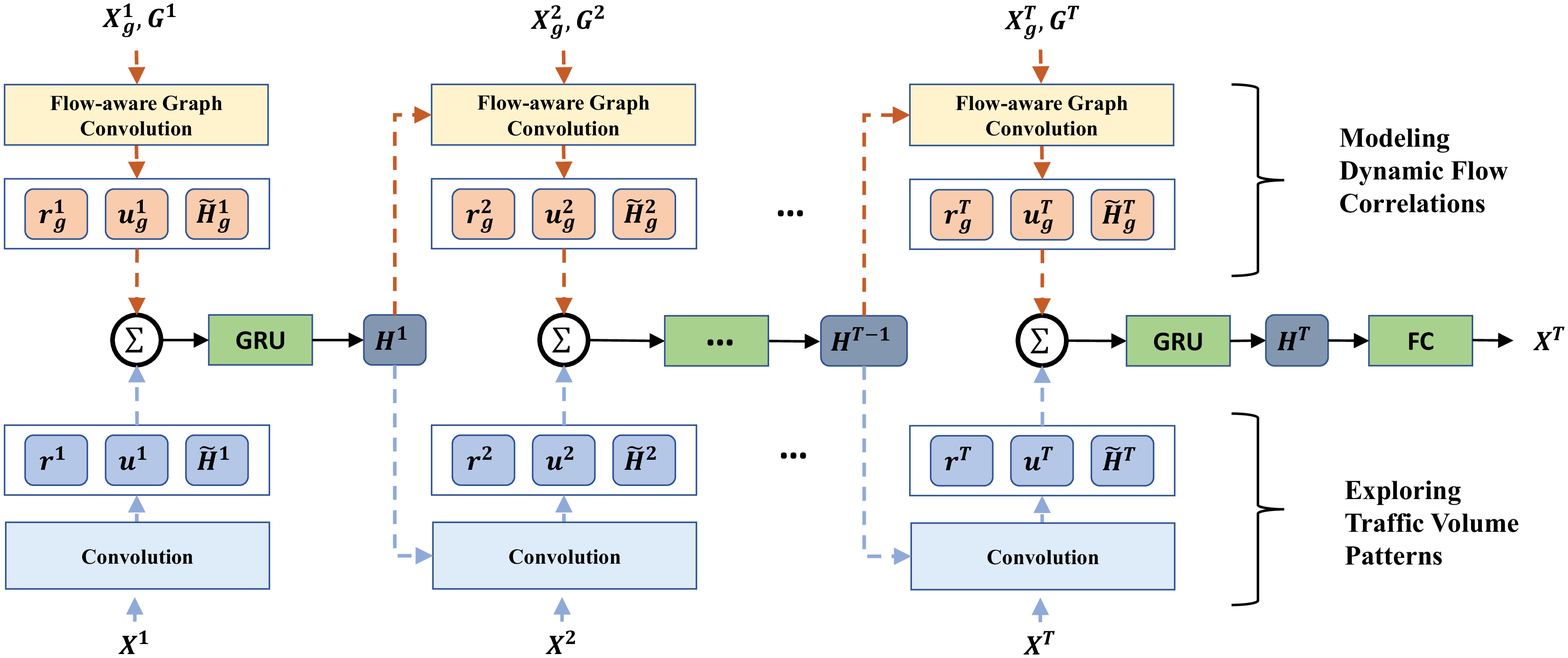}
	\caption{Illustration of Proposed Approach.
	}
	\label{fig:framework}
\end{figure*}
}

\begin{figure}[t]
	\centering
	\includegraphics[width=\linewidth]{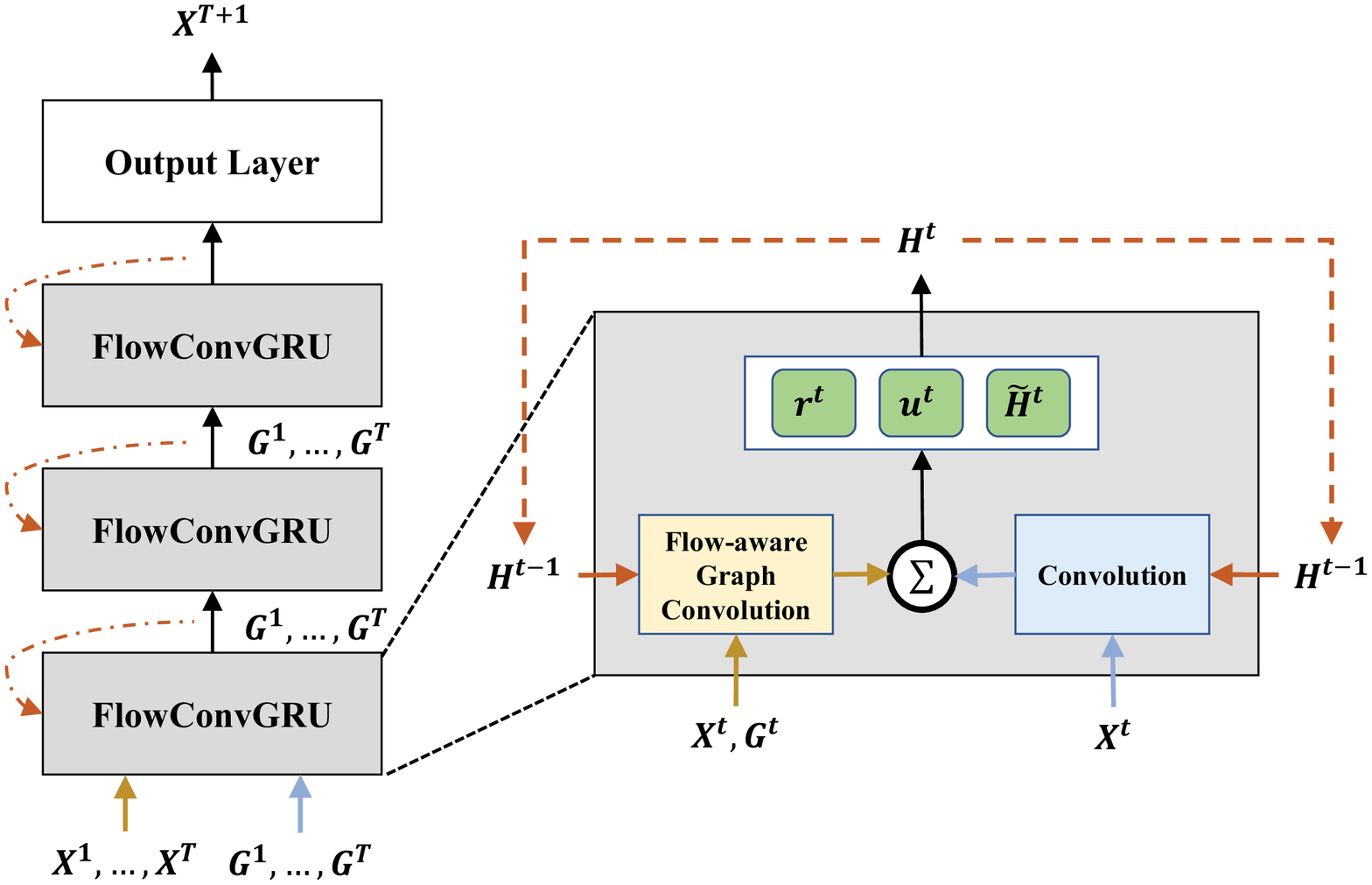}
	\caption{Architecture of our proposed model.
	}
	\label{fig:framework}
	\vspace{-.1in}
\end{figure}

\section{Methodology}

Figure~\ref{fig:framework} provides the architecture of our proposed approach to traffic prediction. 
We consider two kinds of inputs: traffic volume tensors and traffic flow graphs over the previous $T$ time steps.
We develop a flow-aware graph convolution module to capture the correlations among regions based on dynamic traffic flows. 
We then apply typical convolutions over traffic volume tensor to learn spatial correlations for nearby regions. 
To capture the temporal traffic tendency, we leverage the Gated Recurrent Unit (GRU) to absorb the outputs from both convolutions. Considering the deep interactions between traffic volume and traffic flows, we couple their features at each step of the GRU. 
We dub the above integrated structure as \textbf{FlowConvGRU}. In our proposal, a stack of FlowConvGRU is constructed to learn deep spatiotemporal features and interactions, followed by a fully connected layer to predict the traffic volume tensor in time step $T+1$. 
In what follows, we provide the details of each module in our method.

\subsection{Modeling Dynamic Flow Correlations}

The input to this module is the traffic volume tensor $X^t$ and the traffic flow graph $G^t$ at time $t$. 
As described before, the direct traffic flow between two regions indicates the correlation in their traffic volume. A novel attempt of our approach is to apply graph convolution to reveal the flow-aware traffic correlations among regions during t. To do this, we simply reorganize the 3D traffic volume tensor $X^t\in \mathbb{R}^{m\times k\times 2}$ into a graph signal $X_g^t \in \mathbb{R}^{N\times 2}$. 
Note that typical graph convolution requires the existence of a fixed graph structure, while there is no stable or explicit edges among regions. We thus resort to the flow graph $G^t$ to identify the dynamic relationships among regions. 
It is also worth mentioning that the flow information is effective to reveal the direct traffic correlations for far away regions. 

Recall that $f^t$ is the adjacent matrix for $G^t$, containing weighted directed edges among regions. 
We employ diffusion convolution on traffic flow graph $G^t$.
At each time $t$, for a region $i$, all its flow-connected regions (according to $G^t)$ actually composes the receptive field $R^t_i$ of $i$ in the diffusion convolution at $t$, which could also change over time.
We first calculate the in-degree and out-degree diagonal matrices $D_I^{-1}$ and $D_O^{-1}$ for the flow matrix $f^t$ and obtain state transition matrices $D_I^{-1} (f^t)^T$ and $D_O^{-1} f^t$.
We then apply diffusion convolution to all the dimensions of the input graph signal $X_g^t\in \mathbb{R}^{N\times 2}$. Let $Z^t \in \mathbb{R}^{N\times Q}$ denote the output graph signal and $K$ be the diffusion step. Our flow-aware graph convolution is formulated as follows:
\begin{equation}\label{eq:gcn}
Z^t(:, q) = \sum_{p=1}^2 {\Theta_{p,q,:,:}} \star^{f^t}_{\mathcal{D}} X^t(:, p), ~~~~\forall q \in \{1, \cdots, Q\}
\end{equation}
where $\Theta \in R^{2 \times Q \times K \times 2}$ is parameter tensor to be learned;
$\star^{f^t}_{\mathcal{D}}$ denotes the diffusion convolution with state transition matrices calculated by $f^t$.
It is important to notice that the state transition matrices $D_I^{-1} (f^t)^T$ and $D_O^{-1} f^t$ at each time step can be calculated efficiently with the time complexity of ${\rm O}(N+|E^t|) \ll N^2$.

Note that the proposed flow-aware graph convolution can be stacked into a deeper structure for modeling high-order flow correlations. It can be simply achieved by taking the output graph signal as input into another flow-aware graph convolution, using the same traffic flow graph $G^t$.

\begin{equation}
w(t) = \frac{1}{\sqrt{2\pi} \cdot \sigma} e^{\frac{-t^2}{2 \sigma^2}} 
\end{equation}

$\psi(t) = \frac{1}{\sqrt{2 \pi} \cdot \sigma^3} \left( e^{\frac{-t^2}{2 \sigma^2}} \cdot \left( \frac{t^2}{\sigma^2} - 1 \right) \right) 
$

$w(t) = e^{i a t} \cdot e^{-\frac{t^2}{2\sigma}} s = 2^{\, j}$

$y[n] = \sum\limits_{k = -\infty}^\infty h[k] \cdot x[2n - k] 
$

\subsection{Modeling Spatial Correlations}

The input to this module is the traffic volume tensor $X^t$ at time $t$. In practice, there are spatial correlations among the traffic volume in nearby regions. For example, several adjacent regions may belong to the same business center and their traffic volume has similar trend over time. As CNN prioritizes spatial locality, we apply a 2D convolution over $X^t$ to extract spatial features. Note that $X^t$ is analogical to an image with $m\times k$ pixels and two channels. Given a 2D convolutional filter $W$ and $X^t$, the output feature tensor $\tilde{X^t}$ can be formulated as:
\begin{equation}\label{eq:conv}
\tilde{X^t} = W * X^t
\end{equation}
where $*$ denotes the convolution operation. In fact, there can be more than one filters to extract different spatial features, which will lead to an output tensor with multiple channels. Here we present the convolution operator with a single filter for simplicity.


\subsection{Flow-aware Traffic Prediction Network}

Given a sequence of traffic volume tensors $\{X^1, \cdots, X^T\}$ 
and that of traffic flow graphs $\{G^1, \cdots, G^T\}$, 
we apply Equation~(\ref{eq:gcn}) and~(\ref{eq:conv}) at each time step.
Specifically, the two kinds of convolutions mainly capture the flow and spatial correlations among regions within a single time period. In order to predict $X^{T+1}$, we have to learn the temporal tendency of traffic volume from time $t=1$ to $T$.
There are many versions of RNN, which have achieved great performance in sequence modeling tasks. 
In this paper, we adopt the Gated Recurrent Unit (GRU), an effective variant of RNN, due to its advantages of fewer parameters and less training time. 
But our approach is general and can be seamless incorporated with other RNN variants. 
Basically, each GRU cell is associated with a hidden state and two gates.
At the $t$-th time step, the cell absorbs previous hidden state $H^{t-1}$ and the current input to update the cell status.
The access or modification to the cell is controlled by the reset gate $r^t$ and the update gate $u^t$.
To be specific, the reset gate $r^t$ decides how to combine new input information with past information in the cell, while the update gate $u^t$ controls the information flow from past steps to the current step.

A simple way to absorb the respective outputs $\{Z^1,\cdots, Z^T \}$ and $\{\tilde{X^1}, \cdots, \tilde{X^T} \}$ from graph convolution and typical convolution over $T$ time steps is to construct two separate GRUs and learn the individual temporal dependencies. After that, a combination of the last hidden states from two GRUs can be performed to produce the final prediction result. 
However, a limitation of separately modeling the sequences is that 
the potential feature interactions between $\{Z^t\}$ and $\{\tilde{X^t}\}$ are completely ignored, which may hurt the prediction performance. 
Intuitively, traffic flows are deeply coupled with traffic volume. A heavy traffic flow between two regions may not only affect the traffic volume of two regions, but also the traffic in other regions due to an alternative result from route planning devices. 

In order to model deep feature interactions as early as possible, we propose an integrated network with a single GRU to combine the results from two kinds of convolution modules. 
The inputs to the GRU unit at time step $t$ consist of the traffic volume tensor $X^t$ and flow graph $G^t$ with flow matrix $f^t$, as well as the previous hidden state $H^{t-1}$.
We apply the two kinds of convolutions (Equation~\ref{eq:gcn} and~\ref{eq:conv}) over the inputs together with the previous hidden state $H^{t-1}$ of the GRU unit. 
The updating equations for our proposed GRU is formulated as:
{\small
	\begin{align}
	r^t &= \sigma ( \phi( \Theta_{r} \star_\mathcal{D}^{f^t} [X_g^t, \varphi(H^{t-1})] ) + 
	W_{r} \ast [X^t, H^{t-1}] + b_r) \\
	u^t &= \sigma ( \phi( \Theta_{u} \star_\mathcal{D}^{f^t} [X_g^t, \varphi(H^{t-1})] ) + 
	W_{u} \ast [X^t, H^{t-1}] + b_r) \\
	\tilde{H}^t &= tanh ( \phi( \Theta_{h} \star_\mathcal{D}^{f^t} [X_g^t, \varphi(r^t \odot H^{t-1})] ) + \\
	&\ \ \ \ \ \ \ \ \ \ \ \ \ \ \  W_{h} \ast [X^t, r^t \odot H^{t-1}] + b_r) \\
	H^t &= u^t \odot H^{t-1} + (1-u^t) \odot \tilde{H}^t 
	\end{align}}
where $\phi(\cdot)$ and $\varphi(\cdot)$ are reshape functions to reorganize the input into 3D tensor and graph signal, respectively. Note that the flow-aware graph convolution and typical convolution are performed over input-to-state and state-to-state transitions of GRU, respectively. We dub the above integrated GRU structure as {\bf FlowConvGRU}. 
%

%
%
\noindent{\bf Prediction. }
In order to produce the final prediction result $\hat{X}^{T+1}$, we propose to stack multiple FlowConvGRUs to model deep feature interactions, using the same set of flow graphs $\{G^1, \cdots, G^T\}$.
As shown in Figure~\ref{fig:framework}, we apply 3 FlowConvGRU layers to obtain a high-level representation for historical traffic data.
Finally, we supply the output of the last FlowConvGRU involving $T$ hidden states into a fully connected output layer to generate the traffic volume tensor $\hat{X}^{T+1}$. To summarize, the overall equation for our proposed method is as follows:
\begin{equation}
\small
\hat{X}^{T+1} = {\rm MLP}({\rm FlowConvGRU}(...(\{G^t\}\{f^t\}\{X^t\})))
\end{equation}



\eat{
	Our integrated model extends ConvGRU network structure to further embed flow-aware graph convolution structure in recurrent networks.
	Specifically, ConvGRU adopts convolutional operations in calculations of gates and new hidden states to capture spatial information.
	To combine flow correlations, we further add flow-aware graph convolution in both input-to-state and state-to-state transitions.
}

\subsection{Loss Function}

We use the L2 loss function to train and evaluate our proposed model. Given a training example with the ground-truth traffic volume tensor $X^{T+1}$, the loss function is defined as follows:
\begin{equation}
\mathcal{L}(\Theta) = || \hat{X}^{T+1} - X^{T+1} ||_2^2
\end{equation}
where $\Theta$ are all parameters to be learned, including the filters in both kinds of convolutions, and the parameters in GRU and the final fully connected layer.

\section{Experiments}


\subsection{Datasets}
We conduct experiments on two real-world traffic datasets: \textbf{TaxiNYC} and \textbf{TaxiCD}. The details of the two datasets are described as follows:

\begin{itemize}
	\item \textbf{TaxiNYC}\footnote{\url{http://www.nyc.gov/html/tlc/html/about/trip_record_data.shtml}}: 
	TaxiNYC contains 267,953,551 taxi trip records in New York from 1st January 2014 to 30th June 2015. 
	We split the whole city into $20\times 10$ regions and each region is about $1km \times 1km$.
	We set time interval to 1 hour.
	We use the records from 01/01/2014 to 04/30/2015 as training data, while the records in May 2015 and June 2015 are used as validation and test sets, respectively.
	\item \textbf{TaxiCD}\footnote{\url{https://gaia.didichuxing.com}}: 
	TaxiCD contains 7,065,937 taxicab records in Chengdu, China during November 2016.
	The data were collected by from Didi Chuxing, an on-line car-hailing company in China. 
	The city is split into $20\times 20$ regions and the size of each region is $2km \times 2km$.
	The length of time interval is set to 15 minutes.
	We use the data from 11/01/2016 to 11/25/2016 (25 days) for training, and data in the following 2 days is set as validation set. Data in the last 3 days is used as test data.
\end{itemize}
We preprocess both datasets to generate traffic volume tensors and flow graphs as defined in Section~\ref{sec:def} for each time interval.
We adopt Min-Max normalization to transform the traffic volume and flow to [0,1] scale. 

\subsection{Experimental Settings}
We consider the traffic volume and flow data in the previous 6 time intervals to predict traffic volumes of all regions in next time interval.

\noindent {\bf Evaluation Metrics.}
We evaluate our proposed model with two commonly used metrics: rooted mean squared error (RMSE) and the mean absolute error (MAE).

\noindent {\bf Baseline Methods.}
We compare our model with both basic and advanced methods as follows. 
\begin{itemize}
	\item
	\textbf{HA}: It predicts traffic volume during next time interval by averaging the traffic volumes in previous time intervals.
	\item
	\textbf{ARMA}~\cite{box2015time}: It is a widely used method for predicting future values of time series data. 
	\item
	\textbf{VAR}~\cite{lutkepohl2005new}: Vector Auto-Regressive (VAR) is a multivariate model to capture linear interdependencies among regions.
	\item
	\textbf{FC-GRU}: FC-GRU is the vanilla version of GRU which uses multiplication for input-to-state and state-to-state transition. We flatten the 3D traffic volume tensor and take it as input into FC-GRU.
	\item
	\textbf{DMVST-Net}~\cite{yao2018deep}: A multi-view  spatial-temporal network model, which jointly considers spatial and temporal relations using local CNN and LSTM models.
	\item 
	\textbf{STDN}~\cite{yao2019revisiting}: It utilizes a flow gate mechanism to explicitly model dynamic spatial similarity based on local flow information.
\end{itemize}
Two variants of our proposed model \textbf{FlowConvGRU} are also compared to evaluate the effectiveness of each component.
\begin{itemize}
	\item
	\textbf{FlowConvGRU-nc}: This is our proposed model without the typical convolutional operations in GRU units.
	\item
	\textbf{FlowConvGRU-nf}: Similarly we remove flow-aware graph convolutions in FlowConvGRU to evaluate the effectiveness of modeling dynamic flow correlation. 
\end{itemize}

\noindent {\bf Hyper-parameters.}
The Hyper-parameters are chosen based on performances on validation set.
In our integrated model, the kernel size of convolution on traffic volume tensors is set to $3\times 3$ with a stride of 1, and the diffusion step $K$ in flow-aware graph convolution is set to 2.
The size of hidden states in GRU units is set to 64.
The number of layers for extracting spatial features is set to 3 in our proposed model and all the neural network baselines for a fair comparison.
During the training phase, we adopt mini-batch learning strategy with a batch size of 8 and use Adam~\cite{kingma2014adam} optimizer with a learning rate of $2e^{-4}$ for both datasets.
All the neural network based models are implemented using TensorFlow~\cite{abadi2016tensorflow} or Keras~\cite{chollet2015keras}.

\subsection{Model Comparison}

\begin{table}[!pt]
	\caption{Prediction Results on Two Datasets}
	\centering
	\begin{tabular}{|c|c|c|c|c|}
		\hline
		\bf datasets & \multicolumn{2}{c|}{\bf TaxiNYC} & \multicolumn{2}{c|}{\bf TaxiCD} \\
		\hline
		\bf Methods & \bf RMSE  & \bf MAE & \bf RMSE & \bf MAE \\
		\hline
		HA & 47.997 & 14.942 & 5.88 & 1.289 \\
		\hline
		ARMA & 45.547 & 10.336 & 4.145 & 1.086 \\
		\hline
		VAR & 27.076 & 10.450 & 4.235 & 1.085 \\
		\hline
		FC-GRU & 21.337 & 7.888 & 2.976 &  0.911 \\
		\hline
		DMVST-Net & 20.870 & 7.843 & 2.877 & 0.898 \\
		\hline
		STDN & 21.287 & 8.031 & 2.889 & 0.867\\
		\hline
		\hline
		FlowConvGRU-nc & 21.157 & 8.133 & 3.175 & 0.931 \\
		\hline
		FlowConvGRU-nf & 18.518 & 7.243  & 2.830 & 0.864 \\
		\hline
		\bf FlowConvGRU & \bf 17.714 & \bf 6.992 & \bf 2.817 &  \bf 0.835\\
		\hline
	\end{tabular}
	\label{table:overall results}
\end{table}

Table~\ref{table:overall results} shows the overall results of compared methods over two datasets.
The RMSE and MAE in TaxiNYC are all higher than those of TaxiCD on the methods since traffic volume in TaxiNYC is presented of larger magnitude than TaxiCD.
For both datasets, HA and ARMA report worse results on two metrics, which indicates that there exists large variation on traffic flows in continuous time intervals.
Note that VAR gives better outcome on TaxiNYC compared to the results on TaxiCD.
This is because there are more traffic flows among regions on TaxiNYC dataset due to the smaller grid size and larger time interval.
Thus the interdependency among regions on TaxiNYC could be explored to achieve better performances.
Compared to neural network based models, the worse results in basic methods confirm that the complex spatiotemporal dependencies cannot be well captured with simple regressions.

Among several neural network based models, FlowConvGRU achieves the best performances on both metrics, showing 2.1\%$\sim$16.8\% lower RMSE and MAE than DMVST-Net and STDN.
The reasons are two-fold.
First, instead of using the local traffic volume data to explore spatial features in DMVST-Net and STDN,
FlowConvGRU takes the whole traffic volume tensor as input and leverages flow-aware graph convolutions to explore flow correlations in flow-connected regions.
Although dynamic flow information is also utilized in STDN, it still focuses on a local area and commits to model the dynamic spatial similarity for the local regions.
Second, the flow-aware graph convolution dynamically recognizes flow correlations among the regions with real-time traffic flows.
The other regions are automatically ignored while FC-GRU considers all the regions regardless of whether they have traffic flows.
Notice that FlowConvGRU outperforms both variants on two datasets, which confirms both flow correlation and spatial dependency modeling enhance overall prediction performance.
We also find that FlowConvGRU-nc has worse results than FlwConvGRU-nf.
We analyze that applying only flow-aware graph convolution in GRU units may lose the spatial information in traffic volume since traffic volumes in nearby regions are usually related while flow correlations may not exist between them.
Therefore, in our proposed model, we utilize a coupling mechanism to combine the two kinds of convolutions to model both flow correlations and spatial influences. 
%


\subsection{Analysis of Dynamic Traffic Flow}

\begin{figure}[t]
	\centering
	\subfigure[Change of flow connection regions]{
		\label{fig:nyc-jaccard}
		\includegraphics[width=0.47\linewidth]{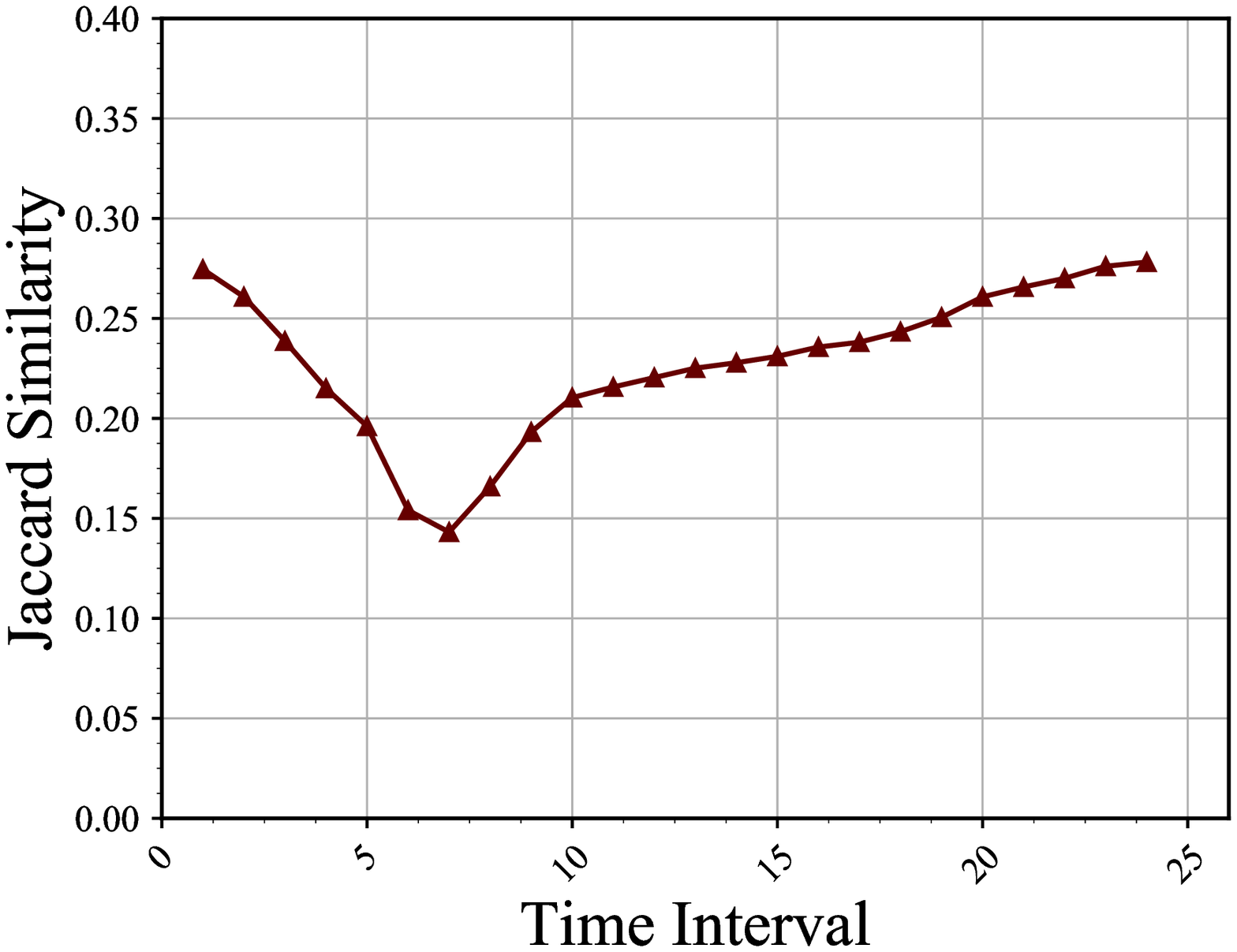}
	}
	\subfigure[Change of traffic flows]{
		\label{fig:nyc-w}
		\includegraphics[width=0.47\linewidth]{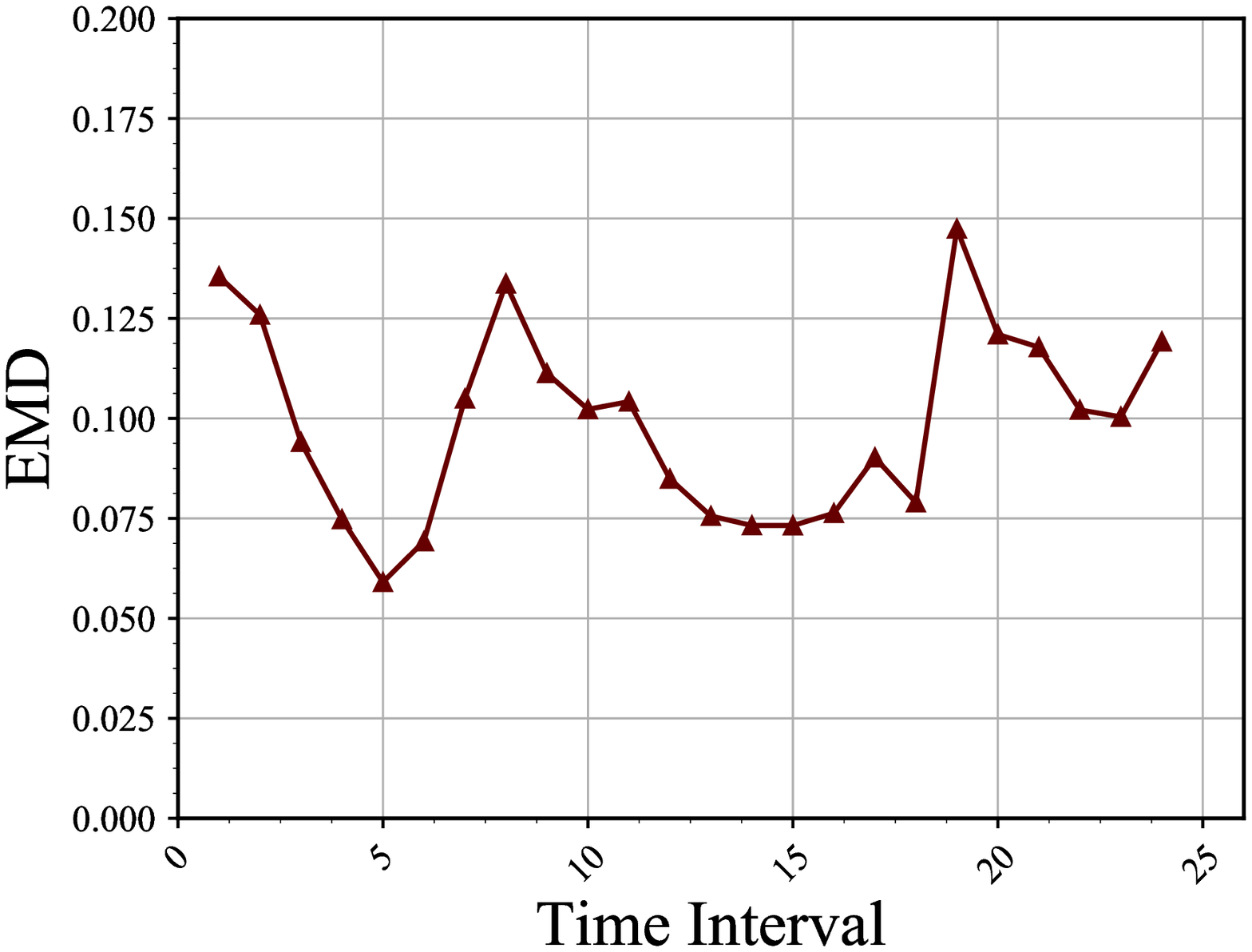}
	}
	\caption{Flow Dynamics in Different Time Intervals}
	\label{fig:flow-analysis}  
\end{figure}

In this section, we present the analysis of dynamic traffic flows among regions and show the advantages of our proposed model in cases with great change of traffic flows.
We use two metrics to analyze temporal dynamics in traffic flows: Jaccard Similarity and Earth Mover's Distance (EMD)~\cite{ijcv/RubnerTG00}.
Suppose any two traffic flow graphs $G^t$ and $G^{t+1}$ in consecutive time intervals. 
For each region $i$, consider the set of connected regions in $G^t$, i.e., the receptive field $R^t_i$ of region $i$ in flow-aware graph convolution at time $t$.
We first compute Jaccard similarity of receptive fields $R_i^t$ and $R_i^{t+1}$ for each region $i$.
Basically a high Jaccard score indicates a larger amount of overlapped connected regions.
And then we average the Jaccard similarity scores of all regions to represent the change of flow-connected regions $\Delta R^t$ in consecutive time $t$ and $t+1$.
Similarly, we calculate the EMD distance between in-flows $f^t_{:i}$ and $f^{t+1}_{:i}$ for each region $i$.
The average EMD distance of all regions is used to represent the change of traffic flows $\Delta f^t$ in consecutive time $t$ and $t+1$.
Figure~\ref{fig:flow-analysis} shows both change of flow-connected regions and traffic flows for 24 different time intervals, i.e., 24 hours in a day.
The Jaccard scores/EMD distances in the figure are calculated by averaging the $\Delta R^t$/$\Delta f^t$ in all corresponding time intervals in TaxiNYC dataset (e.g., all $\Delta R^t$ in 8am and 9am).
We observe that, 
(1) the Jaccard scores are mostly less than 0.25, which means 
at most time there are no more than 25\% regions which will hold the flow connection with same region in next time interval.
(2) there are larger variations of traffic flows during morning and evening peak hours (i.e., 6-8am and 7-9pm with larger EMD distances).

\begin{figure}[t]
	\centering
	\subfigure[TaxiNYC]{
		\label{fig:nyc-large-w}
		\includegraphics[width=0.47\linewidth]{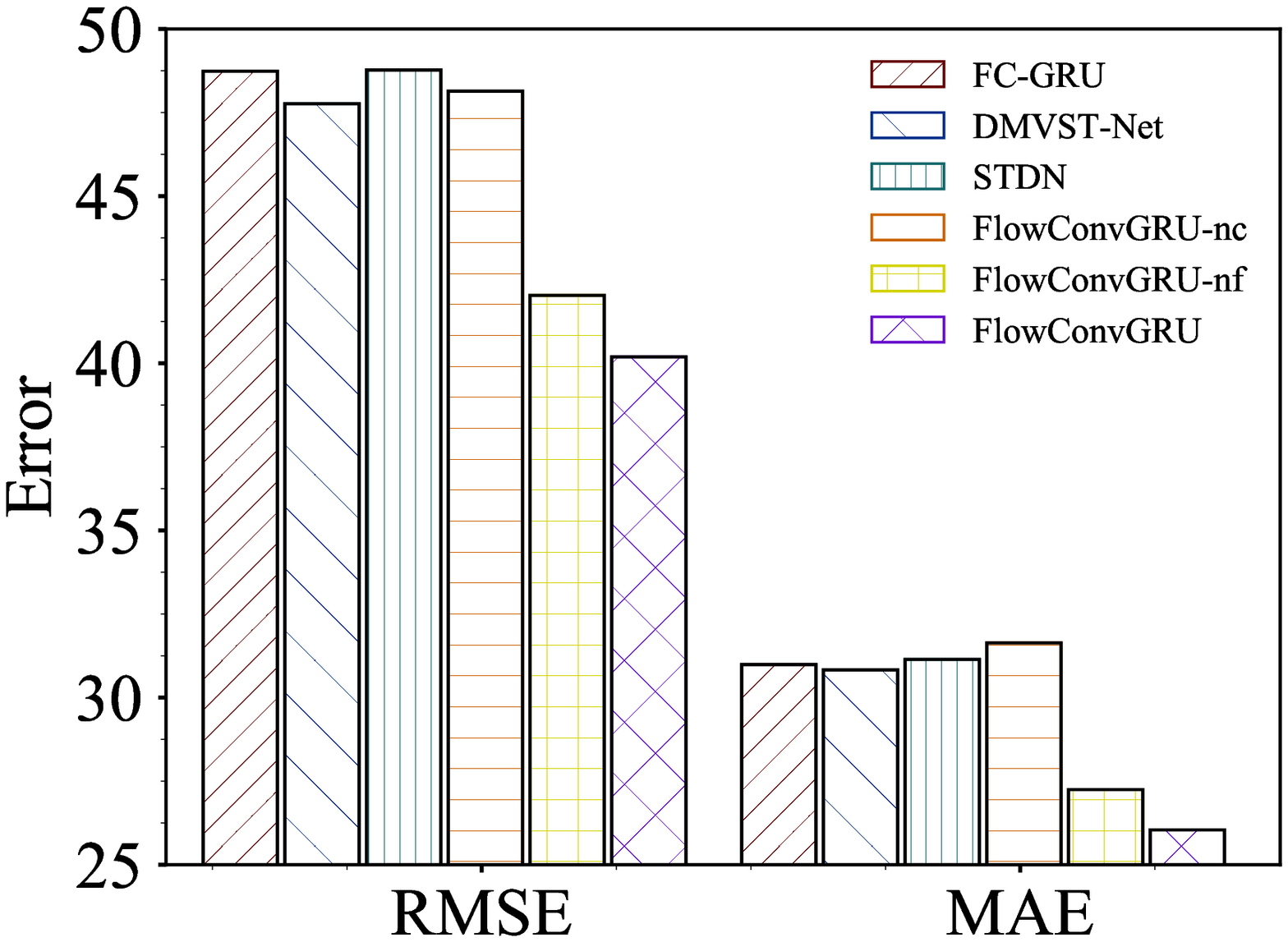}
	}
	\subfigure[TaxiCD]{
		\label{fig:cd-large-w}
		\includegraphics[width=0.47\linewidth]{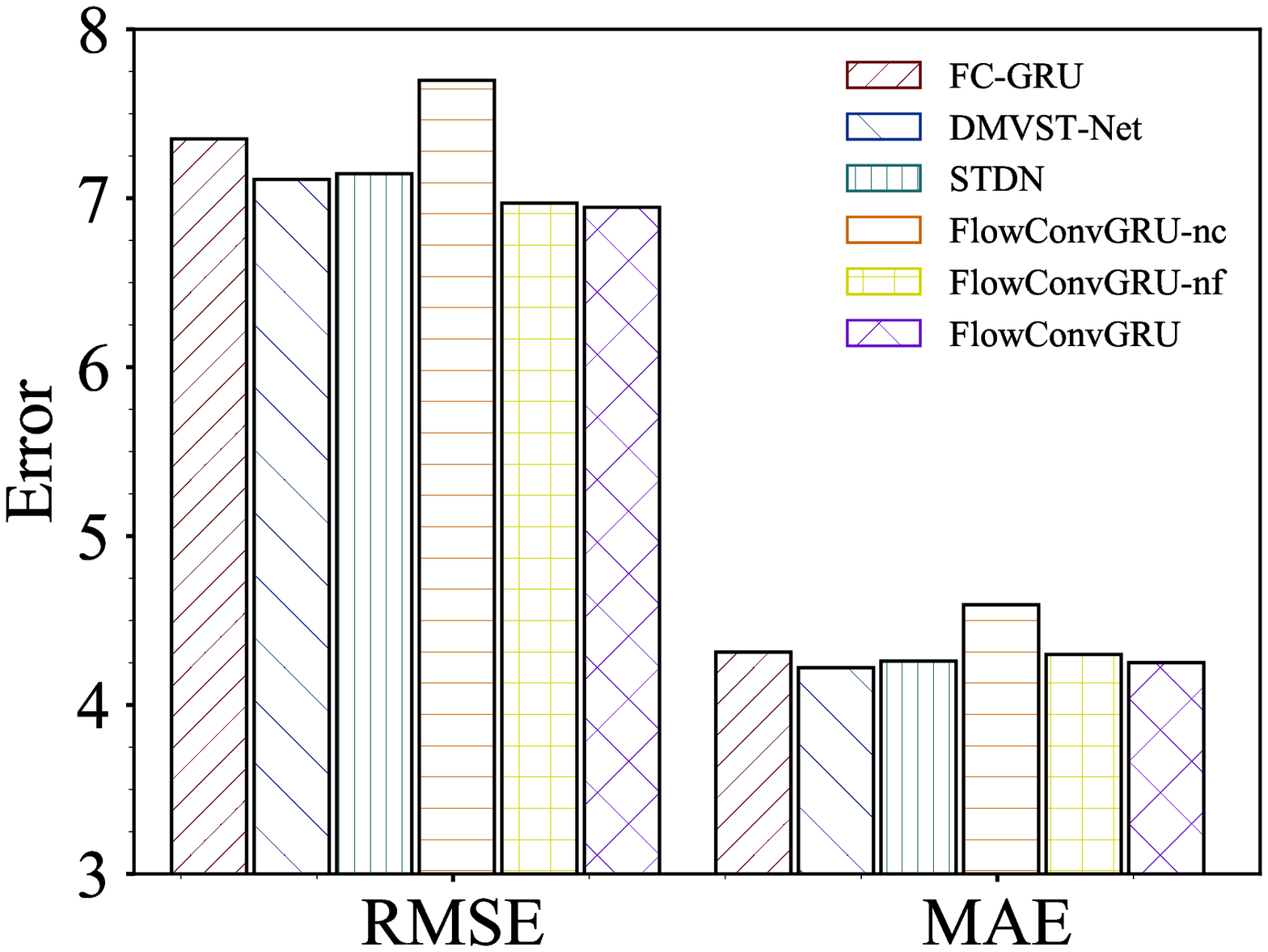}
	}
	\caption{Results on Instances with Great Change of Traffic Flows}
	\label{fig:large_w}  
\end{figure}

To evaluate the traffic prediction performance in conditions with great change of traffic flows, we set a threshold of EMD \textit{w} and evaluate prediction results on instances with EMD distance higher than \textit{w}, i.e., conditions with great change of trafffic flows.
We set the threshold as 0.1 and 0.005 for TaxiNYC and TaxiCD, respectively.
Obviously, instances with EMD distance higher than 0.1 in Figure~\ref{fig:flow-analysis}(b) are from morning and evening peak hours.
The prediction results are provided in Figure~\ref{fig:large_w} (we have omitted the worse results of HA, ARIMA and VAR for simplicity).
We can see that the errors are apparently much higher than overall prediction errors presented in Table~\ref{table:overall results}, which indicates that the conditions with great change of traffic flows are mostly with heavy traffic and difficult to predict.
Nevertheless, our proposed model could also achieve better performance, which 
demonstrates the ability of capturing dynamic flow correlations effectively.

\subsection{Effects of Number of Layers}

\begin{figure}[t]
	\centering
	\subfigure[TaxiNYC]{
		\label{fig:nyc-layer}
		\includegraphics[width=0.47\linewidth]{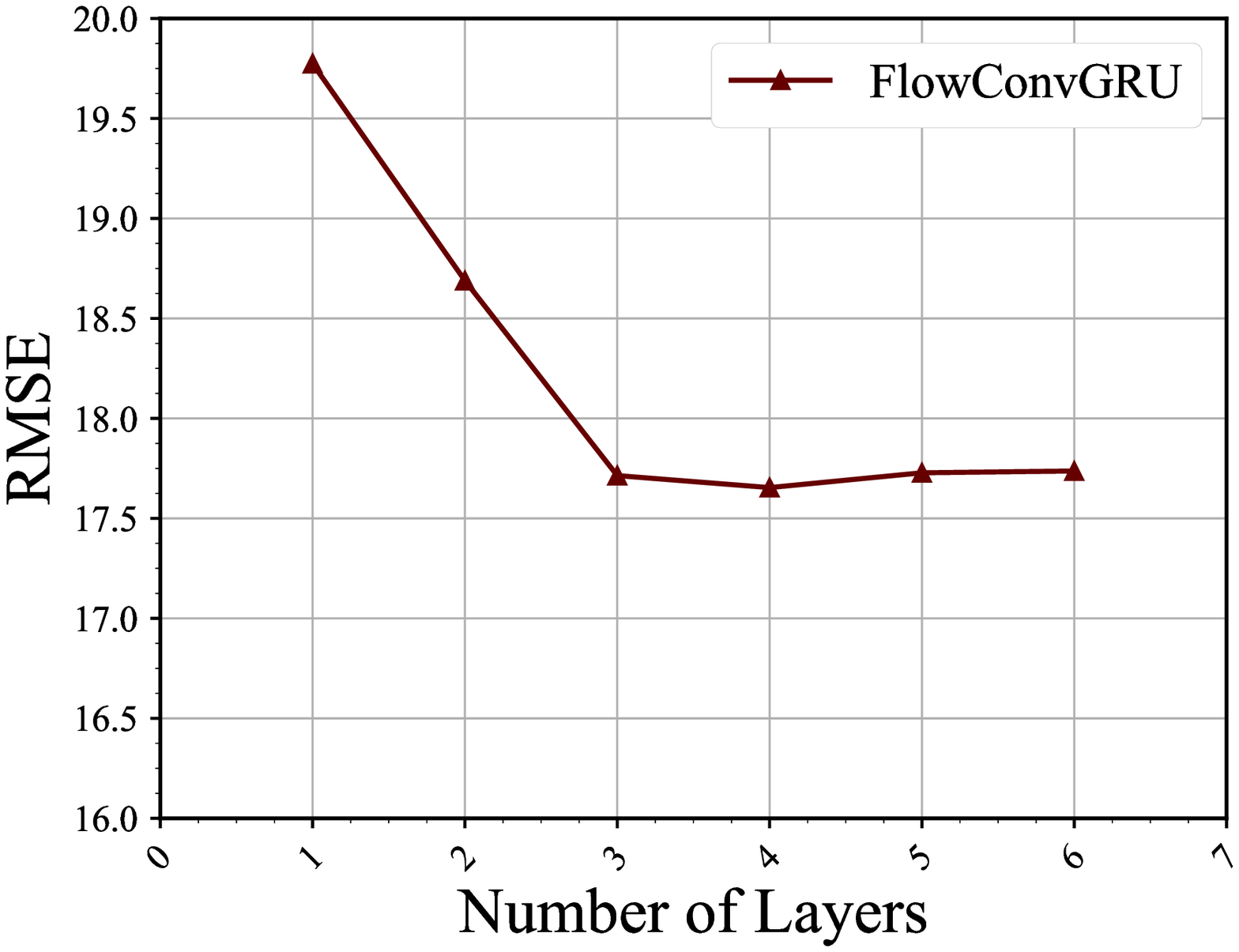}
	}
	\subfigure[TaxiCD]{
		\label{fig:cd-layer}
		\includegraphics[width=0.47\linewidth]{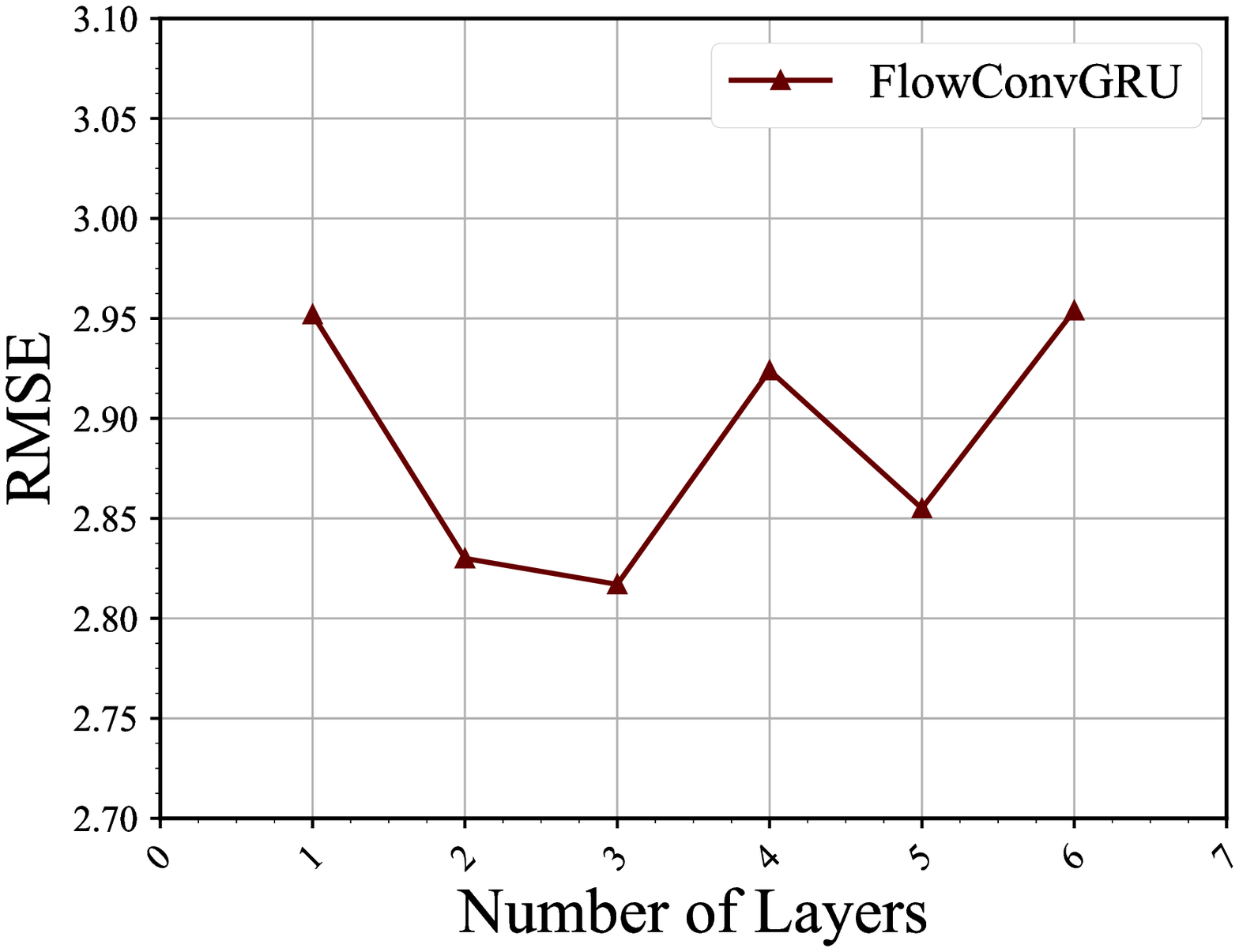}
	}
	\caption{Results on Different Number of Layers}
	\label{fig:num-layer}  
\end{figure}

Figure~\ref{fig:num-layer} presents the prediction results of FlowConvGRU with different number of layers.
For both datasets, 
the best performance is achieved with 3 FlowConvGRU layers.
Shallow network structure may fail to capture spatiotemporal correlations in distant area while deeper models could cause overfitting and lead to higher generalization error.
In fact, the deeper network architecture considers the spatial correlations with farther area and more forward or backward flows.
As the network gets deeper, more spatial information can be utilized but at the same time more unconnected regions will also be taken into account, which may oppositely cause performance degradation.

\section{Conclusion}

In this paper, we introduce a novel flow-aware graph convolution to explore dynamic flow correlations in traffic prediction problem.
We explicitly construct the traffic flows among regions as graph data and apply graph convolutions over the flow graph.
By utilizing real-time traffic flow information, the proposed flow-aware graph convolution is able to disclose dynamic flow correlations among regions, even in distant regions. 
To model both flow and spatial correlations,
we propose an integrated network model based on gated recurrent units to capture spatiotemporal features in nearby regions and those with traffic flows.
Our experiments on two real-world datasets show that the proposed integrated network model outperforms the state-of-the-art methods.
And the integration of flow-aware graph convolution in GRU can effectively improve the accuracy, especially on the cases with great change of traffic flows.

\bibliographystyle{named}
\bibliography{reference}

\end{document}